# Transdisciplinarity seen through Information, Communication, Computation, (Inter-)Action and Cognition

Gordana Dodig-Crnkovic[1], Daniel Kade[2], Markus Wallmyr[2], Tobias Holstein[2] and Alexander Almér[3]

**Abstract**

Similar to oil that acted as a basic raw material and key driving force of industrial society, information acts as a raw material and principal mover of knowledge society in the knowledge production, propagation and application. New developments in information processing and information communication technologies allow increasingly complex and accurate descriptions, representations and models, which are often multi-parameter, multiperspective, multi-level and multidimensional. This leads to the necessity of collaborative work between different domains with corresponding specialist competences, sciences and research traditions. We present several major transdisciplinary unification projects for information and knowledge, which proceed on the descriptive, logical and the level of generative mechanisms. Parallel process of boundary crossing and transdisciplinary activity is going on in the applied domains. Technological artifacts are becoming increasingly complex and their design is strongly user-centered, which brings in not only the function and various technological qualities but also other aspects including esthetic, user experience, ethics and sustainability with social and environmental dimensions. When integrating knowledge from a variety of fields, with contributions from different groups of stakeholders, numerous challenges are met in establishing common view and common course of action. In this context, information is our environment, and informational ecology determines both epistemology and spaces for action. We present some insights into the current state of the art of transdisciplinary theory and practice of information studies and informatics. We depict different facets of transdisciplinarity as we see it from our different research fields that include information studies, computability, human-computer interaction, multi-operating-systems environments and philosophy.

## 1. Introduction

There is no human today who would possess all knowledge of even one single classical research discipline. In the case of physics, Wikipedia, Outline of Physics, lists 30 different branches of which many have several important sub-branches. The exact number can of course be disputed, but it is evident that they are far too many for an individual researcher to know in depth. As the amount of knowledge constantly grows, and the process of forgetting and loosing previous knowledge nearly gets completely extinct, but on the contrary old sources get digitalized and made available, amount of information and knowledge dramatically increases and specialization, branching and division into new sub-disciplines continues. On the other hand a process in the opposite direction of synthesis and increased connectivity is becoming more and more prominent as well. Based on information communication technologies, humanity is becoming networked through variety of interactions and exchanges constantly going on – from information and knowledge, to money and things, objects, goods, commodities. Communication and exchanges create global society with its global and complex problems – from climate change, pollution, question of resources and other environmental issues that threaten sustainable development, to complex social topics of mass migrations, long-term urban planning and healthcare dealing with epidemics prevention or understanding of diseases on multiple levels of organisation, from molecular to organismic level - to name but a few topics. Complex global problems are calling for systemic, both broad and deep understanding. Also the developments of new technologies, such as internet of everything, digitalization of society, autonomous vehicles, industrial and social robotics, intelligent cities, homes, and infrastructures … all may be expected to radically change our civilization, and presuppose decision making and problem solving based on knowledge from many traditionally disparate disciplines that range from natural and technical sciences to humanities and arts. They necessitate a team work which is real-life problem oriented and has high direct societal value that necessitates inclusion of variety of stakeholders – such as governmental, industrial and general public actors. As a response to the demand of complex systems understanding, anticipation of behavior, and control, new synthetic knowledge is constantly developed by fusion and cross-pollination of existing knowledge. Klein in *The Oxford Handbook of Interdisciplinarity* (Frodeman et al. 2010), differentiates between *'endogenous interdisciplinarity'* with focus on the internal theory building between existing academic disciplines (which might be identified as 'interdisciplinarity proper', in contrast to *'endogenous interdisciplinarity'* driven by real-life problems knowledge integration and could be identified with *transdisciplinarity*. Interdisciplinarity in that context presents a tool for transdisciplinarity, which on top of deep interdisciplinary collaboration between academic research fields adds the factor of real-life relevance and stakeholder involvement.

---

[1] Chalmers University of Technology and University of Gothenburg, Sweden
[2] Mälardalen University, Sweden
[3] University of Gothenburg, Sweden

## 2. Transdisciplinarity vs. multidisciplinarity vs. interdisciplinarity

*"So many people today—and even professional scientists—seem to me like someone who has seen thousands of trees but has never seen a forest. A knowledge of the historical and philosophical background gives that kind of independence from prejudices of his generation from which most scientists are suffering. This independence created by philosophical insight is—in my opinion—the mark of distinction between a mere artisan or specialist and a real seeker after truth."*

A. Einstein to R. A. Thornton, unpublished letter dated Dec. 7, 1944; in Einstein Archive, Hebrew University, Jerusalem, as quoted in (Cooper et al. 2007)

Before we start, we briefly introduce some definitions of terms we are going to use. As our focus will be on transdisciplinarity, we just briefly outline the difference between transdisciplinarity, interdisciplinarity and multidisciplinarity.

**Monodisciplinary research**

In our approach we adopt the view of discipline as a part or a subsystem of a bigger architecture of the knowledge production. According to (Choi & Pak 2008), "A discipline is held together by a shared epistemology. (…) The proposed conceptual framework of the knowledge universe consists of several knowledge subsystems, each containing a number of disciplines." Unlike Choi and Pak, we do not see knowledge production in the first place as a hierarchy (even though there is a hierarchy of levels of scale or granularity of domains), but as a network of networks of interrelated disciplinary fields (Dodig-Crnkovic 2016). A discipline corresponds to an academic field of research and education that typically has its own journals and academic departments. Disciplinary research is termed Mode-1 (Nowotny et al. 2001), while Mode-2 stands for the production of knowledge through interdisciplinary and transdisciplinary research close to a context of application.

**Multidisciplinary research** - working *with* several disciplines - implies that researchers from different disciplines work together on a common problem, but from their own disciplinary perspectives. According to the Klein taxonomy, the main characteristics are juxtaposing, sequencing and coordinating of knowledge (Klein 2010).

**Interdisciplinary research** - working *between* several disciplines**,** used to denote the setting where researchers collaborate transferring knowledge from one discipline to another. According to the Klein taxonomy, the main characteristics are integrating, interacting, linking, focusing and blending.

**Transdisciplinary research** – From the meaning of the Latin word *trans,* Nicolescu derives the definition of transdisciplinarity as that knowledge production which is at the same time *between*, *across* and *beyond* all disciplines, (Nicolescu 2014). Transdisciplinarity is a research approach that enables addressing societal problems through collaboration between research disciplines as well as extra-scientific actors. It enables mutual learning among and across disciplines as well a between science and society. The main cognitive challenge of the research process is integration which is based on reflexive attitude both oriented towards different actors in the research process and their mutual relations, and towards the research project as a whole in its context (Jahn et al. 2012) The main difference to interdisciplinarity, apart from the degree of interaction, is the involvement of extra-scientific stakeholders in transdisciplinary research. (Frodeman et al. 2010) (Hadorn et al. 2008) In the course of the research process, boundaries between disciplines dissolve through integrated perspectives, knowledge and approaches from different scientific disciplines and other external sources interfuse, (Flinterman et al. 2001). Transdisciplinarity is often applied to address the real world complex problems through context-specific negotiation of knowledge that emerges from collaboration. (Thompson Klein 1996) Research fields include environmental-, sustainability-, gender-, urban-, cultural-, and peace and conflict-, future-, public health- and information studies, policy sciences, criminology, gerontology, cognitive sciences, information sciences, materials science, artificial intelligence, human-computer interaction, interaction design, ICTs and society studies, etc. From the organizational point of view, "Transdisciplinary research is, in practice, team science. In a transdisciplinary research endeavor, scientists contribute their unique expertise but work entirely outside their own discipline. They strive to understand the complexities of the whole project, rather than one part of it. Transdisciplinary research allows investigators to transcend their own disciplines to inform one another's work, capture complexity, and create new intellectual spaces." (Güvenen 2015) Involvement of stakeholders providing the context for the solution of real-world problems is central for transdisciplinary research. Distinctive characteristics of transdisciplinary research, according to Klein taxonomy are transcending, transgressing and transforming. (Klein 2010)

Finally, It is important to realize that disciplinary, multidisciplinary, interdisciplinary and transdisciplinary research present different forms that complement and presuppose each other and by no means exclude or replace. There is still however a lot of uncertainty and confusion about the meaning of each form of knowledge production and their mutual relationships. With regard to ontological status of transdisciplinary research, Brenner argues "transdisciplinarity should not be seen as yet another discipline but as an aid to legitimizing and insuring a minimum scientific rigor in creative new approaches to on-going issues." (Brenner & Raffl 2011) To this day, interdisciplinarity and transdisciplinarity is hardly seen at universities, and their slow introduction happens indirectly through courses addressing e.g. sustainability. However, there is a strong need of introducing this knowledge broadly and making it part of curricula so that the next generation of researchers get prepared for work in all types of constellations – from mono-disciplinary to transdisciplinary research, being

"vaccinated" against disciplinary chauvinism. The aim of our article is to contribute to better understanding of the existing knowledge production practice and theory based on research connecting information communication, computation, (inter-)action and cognition.

## 3. Diversity of Sciences, Humanities and Arts

Sciences as we know them today are historically new phenomenon. Since the dawn of western civilization with Aristotle in the ancient Greece and to the 19th century all study of the natural world was known as natural philosophy. Newton, Lord Kelvin, Spinoza, Goethe, Hegel and Schelling were natural philosophers. With the development of specific natural sciences like astronomy, physics, chemistry, biology etc. natural philosophy faded into near nonexistence. Other two branches of traditional philosophy, metaphysics and moral philosophy, continued to this day to study fundamental nature of reality, knowledge, reason, mind, language and values. They contributed and were integrated into the development of humanities and arts and help us get a broader understanding of human conditions. The development of early sciences proceeded by replacing the question "why" (is something)? i.e. the question of *telos*, or the purpose, goal of something, with the question "*how*" (is something possible)? or in what way exactly it happens – that is still the focus of modern sciences. Especially *the question "why?"* as related to Aristotle's final cause was strictly exorcised from modern science, such as Galilean and Newtonian physics, based on classical (linear, exact) logic.

However, in last decades it has become apparent that Aristotle's teleological processes could be described and scientifically modeled with help of memory-based agency such as in living organisms. All living organisms must actively "work" on their own survival – without appropriate environment, food and water an organism cannot exist. That makes them sensitive to the environment where they anticipate future possibilities: they avoid dangers and choose favorable circumstances. Organisms anticipate probabilistically, based on memory of previous experiences. From the contemporary perspective, Aristotle's final cause is nothing mystical, as it is the result of living beings survival strategies – it is not based on an exact knowledge of the future, but on the probabilistic expectation and anticipation. Among living organisms, humans have developed the most sophisticated strategies of anticipation based on learning that is both individual and collective/cultural.

Kant argues in his three Critiques (the Critique of Pure Reason, the Critique of Practical Reason, and the Critique of the Power of Judgment) that all human understanding which shapes our experience is *teleological*, i.e. *goal oriented*. He introduces *judgment* as a basis of decision and action and a way to unify the theoretical and practical perspective (Hanna 2014). Typically we know something because we find it relevant, important and interesting, and useful for acting in the world. All of it is based on *values and judgment*: what we find good and worthy of our time and efforts. Knowledge and values are inextricably connected (Tuana 2015).

With current prominence of problem-based research, development of increasingly complex technology and taking into account variety of stakeholders involved – the necessity for a broader understanding by each of participants in such projects is becoming central. We again need to acquire a broader view in which sciences, humanities, arts and other human activities form networks of networks of tightly interrelated parts, as we are becoming aware of the complexity of the natural and cultural worlds and ready to approach it. (Bardzell & Bardzell 2015)

Natural sciences (primarily physics with its fields of mechanics, thermodynamics and electromagnetism) were the basis for the development of technology that has led to the modern industrial era. Mechanistic ideal of physics have permeated other fields and its strict division of labor appeared for centuries as natural necessity. Even bigger was the division between natural sciences and humanities and arts. Almost sixty years ago, Snow famously addressed the gap between natural sciences and humanities in his book *The Two Cultures* (Snow 1959). A manifestation of a deep schism, in the 1990s science wars raged between scientific realists and postmodernists, epitomized by Sokal affair, (Sokal & Bricmont 1997). The starting point was Sokal's hoax article "*Transgressing the Boundaries: Towards a Transformative Hermeneutics of Quantum Gravity*" which was caricaturing the relationships between postmodernism of humanities and realism of natural sciences. As an attempt to bridge the gap, biologist Wilson wrote a book *Consilience: The Unity of Knowledge*, trying to reconciliate "the two cultures" in the academic debate (Wilson 1998). Wilson's proposed solution was "the third culture", which would foster deeper understanding between humanities and natural sciences. Interestingly, the German term Wissenschaft includes both natural and social sciences as well as the humanities, unlike the English concept of "science" that makes the distinction between sciences and humanities. In terms of education, there is a "third way" where liberal arts education can include languages, literature, art history, philosophy, psychology, history, mathematics, and sciences such as biological and social sciences.

However, Snow's model of knowledge production might have worked for a few individuals, but culture is a mass phenomenon and calls for public involvement. Thus the third culture instead started to emerge as a result of technological development, ICT-revolution and digitalization of society in virtually all its segments. (Kelly 1998)(Brockman 1996) Computational devices made it possible to visualize, simulate, communicate and discuss ideas that were before completely inaccessible to the broader audience.

The key for the new knowledge production capable of bridging variety of gaps was in the dialog, collaboration, and crowdsourcing. Such examples are "polymath" online crowd-based mathematical problem solving, and Wikipedia, which shows that crowdsourcing style of public knowledge production can work remarkably well. Maybe the most radical novelty of transdisciplinary research is involving ordinary non-scientific people in the co-production of knowledge together with scientists. It is good to remember, as Nicolescu (Nicolescu 2011) reminds us that given more than 8000 disciplines we can be experts in one but remain equally ignorant as any other common person in all the other thousands of disciplines. Typical real-life problems are complex, often "wicked", and demand expertize in variety of research fields as well as knowledge by acquaintance, experiential knowledge and involvement and engagement in their solution. Examples of such wicked problems are global warming, public health issues or mass migrations. It is necessary to understand kinds of knowledge and skills necessary in addressing of such problems and the process of collaboration and common knowledge production.

Today we have many gaps, big and small between different disciplines. In this article we will argue that the question is not only *how to understand* the world but also how to make decisions and *how to act*. So in what follows we will also indicate the connection between *understanding* and *acting* in our different research projects that built on transdisciplinarity. To start with, we present various projects of unification and synthetic approaches to information and knowledge, which differ in their goals and preferences, focus and applicability.

## 4. Unity through diversity of information processes and knowledge production. Transdisciplinary integration projects

The traditional linear notion of knowledge pictured as a tree that grows only in one direction, from the root to the branches, is today replaced by images of fractal structure, as Klein pointed out in (Klein 2004) or an organic growing rhizome such as in Deleuze (Deleuze & Guattari 2005). More than anything else, we would say, knowledge production today is associated with network of networks that unites fractals with organically growing structures. (Barabasi 2007) (Dodig-Crnkovic & Giovagnoli 2013) Importantly, digital space enables non-linear dialog where information flows in all directions and distributed learning happens in all nodes. Not only so that the central node (such as university or research institute) emits knowledge to the crowds, but crowds more and more actively contribute in knowledge production – as a source of data, opinions, values, preferences and all sorts of other knowledge that might be useful for both problem identification, solution and new knowledge generation.

At present we meet variety of notions of information that focus on one segment, dimension or level of reality, most often without exactly positioning itself in relation to the other existing approaches, frameworks or definitions. In those cases where such presentation is given, it has a form of argument why one's own approach is better (for the chosen purpose) than the others. No attempt is made to pragmatically examine under which circumstances some other approaches, frameworks or definitions would be more appropriate. Thus many unification attempts have been done on different grounds in search for the universal idea of information that would suit all its many appearances and applications.

**Burgin's unified general theory of information (GTI)**

If we want to understand the process of unification of knowledge, the first step is the unification of information. In his book Theory of Information: Fundamentality, Diversity and Unification, (Burgin 2010), Burgin both presents the current state of art in the information studies (Burgin 2010) addressing the most important theories of information such as dynamic, pragmatic, algorithmic, statistical and semantics, as well as presenting his own proposal for the general theory of information (GTI). Burgin's unification is based on a parametric definition of information that uses "infological system" type (infological as information ontological) as a parameter that distinguishes between kinds of information, such as chemical, biological, genetic, cognitive, personal and social, and in that way constructs the general concept of "information". Burgin's general theory of information is a system of principles and there are two groups of such principles: *ontological* (defines information that exists: in nature, in living world including human mind, in societies, even in computing machinery with their "virtual reality", and *axiological* principles that explain evaluation and measurement of information. GTI explicates the relationships between data, information, and knowledge within common framework. With respect to its goals and values Burgin's GTI is a pragmatic and encyclopedic work that aims at logical organisation of information and knowledge with focus on their unity in reality.

**Hofkirchner's unified theory of information (UTI)**

In a different project for unification of information Hofkirchner characterizes efforts at unification into unity of methods, unity of reality and unity of practice, (Hofkirchner 1999).

Starting from the observation that information presents a conceptual building block as fundamental as matter/energy, Hofkirchner argues for the necessity of unified theory of information, UTI conceived as a transdisciplinary evolutionary framework.

"UTI may thus be regarded as a specific proposal of what theoretical foundations of a new science of information could look like, and tries to connect complex systems thinking to systems philosophy and extend it to the field of information studies." (Hofkirchner 2010)

Constitution of sense in this framework is envisaged as three-level architecture of self-organization: cognition, communication and cooperation levels. Different definitions of information correspond to different domains of applicability, and "none of the various existing information concepts/theories should take its perspective absolute but, in a way, complementary to the other perspectives. " Nice illustration is given by Riegler: "Suppose that we take a piece of chalk and write on the blackboard "A = A." We may now point at it and ask, "What is this?" Most likely we will get one of the following answers. (a) White lines on a black background; (b) An arrangement of molecules of chalk; (c) Three signs; (d) The law of identity." (Riegler 2005)According to Hofkirchner, information depends on how we see the object-subject relation.  In "hard" sciences, information is objective, while in "soft" sciences it is subjective. UTI is an integrative framework that aims at bridging the gap. It offers the solution to the Capurro's Trilemma which assumes that the solution to the unification of the concept of information either goes via synonymy, analogy or equivocation (Capurro et al. 1997). The UTI solution is the fourth option – synthetic or integrative approach. (Hofkirchner 2009) This unification does not result in a monolithic body of knowledge but seeks unity through diversity (systemic integrativism). UTI adopts Praxio-Onto-Epistemology: methods of systems philosophy, philosophy of information, social philosophy, philosophy of technology and applying of system methods, evolutionary systems theory and Science of information methods. With respect to its goals and values Hofkirchner's UTI is interested in connecting information with cognition, communication and cooperation in a systemic framework.

**Brier's Cybersemiotics**

Unlike Burgin and Hofkirchner who have information in the focus, Brier is in the first place addressing knowledge, and he declares in his book *Cybersemiotics: why information is not enough!* that information (understood in Shannon's formulation) lacks meaning that is fundamental for living organisms. (Brier 2008) Cybersemiotic is thus used as a "new foundation for transdisciplinary theory of information, cognition, meaningful communication and the interaction between nature and culture". (Brier 2013) According to Brier, phenomenological and hermeneutical approaches are necessary in order to build a theory of signification and interpretative meaning, so he questions the possibility of phenomenological computation, such as proposed by the adherents of info-computation and computing nature (Brier 2014). According to Brier, the bridge from physical information to phenomenology requires metaphysical framework and goes through the following five organizational levels: 1. The quantum physical (information) level with entangled causation. 2. The classical physical (information) level with efficient causation based on energy and force. 3. The chemical informational level with formal causation by pattern fitting. 4. The biological *semiotic* level with non-conscious final causation (where meaning occurs) and 5. The social-linguistic level of self-consciousness, with conscious goal-oriented final causation. Brier argues that integration of these levels made by evolutionary theory through emergent properties is not sufficient, as it lacks a "theory of lived meaning". Cybersemiotics that is offered as a solution for bridging the gap is based on Peirce's semiotic philosophy combined with a Luhmann's cybernetic and systemic view. (Brier 2003) Regarding goals and values,

Briers approach is much more interested in individual subjective information with roots in phenomenological and hermeneutical tradition.

**Integration through qualitative complexity: ecology and cognitive processes**

One more important example of a transdisciplinary unification project is done in the domain of qualitative complexity as described by Smith and Jenks in their book *Qualitative Complexity: Ecology, Cognitive Processes and the Re-emergence of Structures in Post-Humanist Social Theory.* (Smith & Jenks 2006) Their book can be seen as a direct answer to *The Two Cultures* (Snow 1959) with its call for unity of knowledge. Smith and Jenks show the way to move beyond the classical irreconcilable dichotomies (with classical logic of excluded middle) that leave intractable gaps between nature and culture, structure and agency as well as between human and technology. They show how connections can be made between 'humanist paradigm' with its emphasis on human traditional notion of "true knowledge" understood as absolute certainty, and empirically observed oscillations between regularity and contingency, order and disorder in the world. Their unification relies on the insight that humans as well as social systems are special cases of a variety of forms of complex systems. As other complex systems, they are best studied by cross-disciplinary and trans-disciplinary methods. It is a long way ahead before we reach unification, and work out all the details of how complex systems produce culture from nature and agency from structure and back. Smith and Jenks present complexity theory based on conceptual tools from thermodynamics, biology and cybernetics, and explore the emergent and probabilistic aspects of self-organizing phenomena, such as human bodies, ant colonies or markets. 'We are at the beginning […] of a multi-dimensional reunification.' (Smith & Jenks 2006) (p. 276), Complexity theory as an explanatory framework supports a *non-linear and interactive concept of causality*, where small causes can lead to large emergent outcomes. An available energy 'informs' every entity from the non-living to cells, from humans to technological assemblages (p. 243). Complexity provides a very productive framework for exploring dynamic interactions of components interacting in emergent ways in social, natural and technological phenomena described by self-organisation starting with 'a common ontology of matter and information' (p. 95). This approach has a goal to bridge the gap between "two cultures" and build a new "third culture" that connects the two.

**Info-computational synthesis through dynamic networking**

While Smith and Jenks approach has its focus on the bridging the gap between the social and the natural, there is an even bigger project that aims at bridging the gap all the way from the microcosm to macrocosm and back, through all immediate emergent phenomena. It aims at generating knowledge in a variety of domains starting with the most fundamental principles of physics and producing more and more complex. We find such a grand project in Wolfram's New Kind of Science (Wolfram 2002) and trace its idea back to 1676 Leibniz quest for Characteristica Universalis, (Leibniz 1966) a universal language that would define the basis for all knowledge.

Leibniz idea of universal language was related to a Calculus Ratiocinator as a method for generation of true statements via logical calculation that is derivation from common premises, with a plan for a universal encyclopedia that would contain all human knowledge. Leibniz's idea further developed within Hilbert's program of formalization of mathematics, logic and parts of physics. Especially through the development of computing machinery used for processing, storage and communication of information, Leibniz's dream of common language of reasoning started to take concrete and practical forms.

One step further, we can imagine that not only rational reasoning that can be articulated as some sort of language and further on expressed computationally, but the whole of human cognition, including emotions and entirety of embodied human behavior as well can be seen as computational in nature (von Haugwitz et al. 2015)(Dodig-Crnkovic & Stuart 2007) (Dodig-Crnkovic & Müller 2011) (Dodig-Crnkovic & Burgin 2011) (Dodig-Crnkovic 2006). In that case computation is not only logical symbol manipulation but also includes variety of physical, chemical and biological processes going on in human body and its mind (Burgin & Dodig-Crnkovic 2015). Human logical reasoning with symbol manipulation is just a small subset of all natural processes that are going on in humans and that can be described and understood as natural computation, (Ehresmann 2014). Generalizing from Leibniz's project of Characteristica Universalis, we can see not only humans, but also all natural and cultural phenomena, indeed, *the whole of our reality* as manifestations of a variety of computational phenomena. That view is called computationalism, natural computation or computing nature, (Zenil 2012)(Dodig-Crnkovic & Giovagnoli 2013).

Info-computation is a constructive theoretical framework that connects information as a structure and computation as information processing, developed in (Dodig-Crnkovic 2006)(Dodig-Crnkovic 2011)(Dodig-Crnkovic 2009). It synthesizes two approaches: informational structural realism (Sayre 1976) (Wheeler 1990)(Floridi 2003)(Burgin 2010) in which the world/reality is a complex fabric of informational structures, and natural computationalism (Zuse 1970)(Fredkin 1992)(Wolfram 2002) (Chaitin 2007), which argues that the universe is a computational network of networks. Computation is thereby understood in its most general form as natural dynamics, from computational processes in quantum physics, to self-organizing, self-sustaining phenomena such as living organisms and eco-systems. In short, it is continuation and generalization of the same Leibnizian tradition that aimed at common understanding of human behavior, now not only logical reasoning, but its entirety, including human biological, cognitive and social behaviors. Providing mechanisms based on natural computation, from physics, via chemistry to emergent biology and cognition, the info-computational framework enables understanding of mechanisms of science on both object level and meta-levels (that is understanding of understanding). Object level in a sense of describing different phenomena within sciences such as physics, chemistry, biology, neuroscience, etc. as manifestations of the same sort of info-computational structures and processes. Meta levels represent understanding of working mechanisms of cognition and knowledge generation as computation in the info-computational conceptual space.

The proposed unification of sciences in knowledge production diversity goes thus via common language and computational apparatus, and info-computationalism (Dodig-Crnkovic 2010) provides both, in the spirit of Leibniz. Besides classical scientific modeling approaches, it offers additional explorative devices such as simulation, virtual reality and generative models, which Wolfram named "a new kind of science". Emerging info-computational tools such as internet of things and internet of everything offer new means of understanding of the role of human embodiment and embeddedness for production of knowledge through interaction with the physical environment. Info-computing is a method with a capacity of providing a perspective connecting presently disparate fields into a new unified framework comprising natural phenomena from elementary particles to cognitive agents, ecological and social systems, from rational and emotive cognition to (inter)acting in the world. Of special interest is the role of embodied exploratory activity in relation to the virtual and simulated in human-computer interaction, (von Haugwitz & Dodig-Crnkovic 2015).

## 5. Transdisciplinary work in technological applications of information and computation

All the above-mentioned approaches (GTI, UTI, Cybersemiotics, qualitative complexity and info-computationalism) to the topic of transdisciplinary knowledge unification are theoretical in nature. Even though all emphasize in one way or the other the importance of pragmatics, agency, embodiment and embeddedness, they do so on the level of description. Info-computational approaches open however up for a computational language application that directly can connect to physical world through computing systems controlled by programming languages and other info-computational structures and processes. Thus the bridge between the code and the execution, the language and action, the description and practice is made.

The practical involvement with the physical reality meets open contexts of individual and particular, through interaction. Thus design and construction of the physical devices requires transdisciplinarity, which gets implemented and tested, evaluated and reinforced through the research process. We present three examples of transdisciplinary research projects where not only gaps between academic knowledge domains are bridged, but even gaps between theoretical and practical knowledge with its aspects of usability, esthetics and other properties of embodiment and embeddedness.

### 5.1 Transdisciplinarity as a tool for interdisciplinary teams in research and applications in HCI. Experiences from creating a head-mounted display

As a consequence of increasingly complex products and services, and increasingly human-centric, stakeholders-aware understanding of technological artifacts, interdisciplinary teams have become a common trend within research and technology related to engineering companies. Companies have discovered that there is an added value in bringing together in the design of their products knowledge from variety of professions and areas. In particular, the driving force in this trend was the shift in how consumers and society think about products. Products do not only need to be functional but also look and feel good, be sustainable from environmental, social, and economical point of view, ethically produced and used. For these reasons, researchers look into how collaborations between different specializations in research and industry can improve their competences to tackle these issues. We will specifically focus on the relationship between interaction design researchers and engineers.

Combining knowledge from different professions to collaboratively solve a common problem or to develop products is necessary when products and solutions need to be developed that are supposed to be innovative and user friendly. In such circumstances interdisciplinary teams are needed that could collaborate to achieve a better outcome by combining their knowledge. However, this new kind of collaboration is not easy to achieve. Only distributing work and accommodating discussions does not allow for an effective interdisciplinary or transdisciplinary team. It has been shown that interdisciplinarity or transdisciplinarity must be learned and governed (Gray 2008)(Younglove-Webb et al. 1999)(Young 2000). The leader of an interdisciplinary or transdisciplinary team needs to moderate problems with team members, make decisions and suggest methods to achieve common goals. In research it has been noticed that there are challenges when building and steering a successful transdisciplinary team, so leaders with the skills to manage collaboratively may make the difference between success and failure (Gray 2008)

When looking at projects run in industry that are interdisciplinary or transdisciplinary, the practical differences in terminologies seem to converge as it gets harder to distinguish at what point a team was or is interdisciplinary or transdisciplinary. The difference might even get smaller when one team member has transdisciplinary knowledge and uses methods to generate knowledge in a transdisciplinary way.

A challenge for interdisciplinary teams can, as we experienced in our project (Kade 2014; Kade et al. 2015) already start with the task description and its terminology, or even terminologies in general. Other researchers framed this by saying that "working on ill-structured problems or problems with multiple weak dimensions requires more demanding information activities" (Palmer 2006). In our example, a product designer, an engineer and a computer scientist collaborated on creating a head-mounted display (HMD) as a new research prototype. The task was to create a modern HMD with a modern, neat user-friendly design that might be developed into a commercial product.

The task description was rather vague and led to issues that needed to be overcome when the interdisciplinary team started to work together. The engineer directly thought about the latest and greatest hardware to provide up-to-date features. "Modern", "neat" and "design" were interpreted in terms of hardware design and what technical features the HMD should have. The computer scientist was considering the latest software, its structures and how components could communicate in a smooth way. At the same time the designer was wondering what the words "modern", user-friendly and "neat" could mean in terms of a visual design. Some discussions to clarify what should be done did not solve this issue; only more questions on what would be needed arose. After a long first meeting between the three participants, only the computer scientist and the engineer came to a better understanding of what would be needed to build the HMD, as their technical understanding was closer to each other in terms of technical problem solving.

This is only one example showing that when working in an interdisciplinary team, a common language and understanding of the task or problem at hand is of importance. Others already stated, "in transdisciplinary projects, misunderstanding and disagreement are much more likely" (Gray 2008) p.125. Therefore, they need to be resolved or avoided through good team management and work structures. Even the communication between interdisciplinary or transdisciplinary team members would benefit from a common language and understanding.

Therefore, it is important to use adequate terms and descriptions in such a work setup. In our example it would have been better, to give each of the team members a separate task description in the language they are familiar with. For the engineer and the computer scientist, a requirement list and a description of demanded hardware and software components and features with technical terms would have allowed them to understand the starting point and what work was required from their side. The product designer would have been more driven, when the terms "modern" and "neat" would have been enriched with some more details such as "slim", "simplistic" design with an "organic" and "head-band like" form, allowing for an "ergonomic" and "comfortable fit". However, sometimes this level of detail is negotiated between team members without mediation of a leader who would impose decisions in a top-down manner.

Besides questions of language, and communication in general, an interdisciplinary team needs to be lead and governed well to be effective. This means that an effective interdisciplinary team needs a skilled manager with a good understanding not only of the different fields, members of the team belong to, but also of the dynamics, ways of working and resources in the team.

A question in our concrete research project (Kade et al. 2015) was: who should lead the team with the task of creating a new head-mounted display? The designer, the engineer or the computer scientist, or maybe a forth person, trained in project management? It might be any of the above. Apart from the task of managing the project and establishing work and communication flows; issues, misinterpretations and misunderstandings resulting from the interdisciplinary setup of the team must be avoided, identified and solved early on. Nonetheless, it is not self-evident if a project management of an interdisciplinary team should rely on the manager's personality and managing skills or if it needs to be transdisciplinarity skills of a manager who coordinates an interdisciplinary team.

We pose the question if it would for example help to have engineers that understand the work processes and language of a product designer. In our example, this would have improved the situation, as the communication problems or misunderstandings could have been avoided. Managing skills and a suitable personality are certainly of importance to lead a successful interdisciplinary team but the leaders of such teams need to go beyond their field of expertise and should have an understanding of the involved professions, their work terms and ways of working. A good coordination of the work and interactions between the different fields is needed to lead interdisciplinary teams. At the present stage, it is rare that researchers acquire basic competences in interdisciplinary or transdisciplinary research, and new thinking in research education is necessary to remedy this deficiency, as the future of research is in collaboration across disciplinary field borders. Our hope is that next generations of researchers, educated in transdisciplinary thinking will be better prepared to listen and learn and look critically at their own disciplinary knowledge in different contexts and in relation to other fields and disciplines.

When leading interdisciplinary teams, or teams in general, a well-managed distribution of work is a key factor of success of an efficient team. In our specific project, we were interested in the question where do designers stop their work and where do engineers take over. In general, the answer to this question might depend on the setup of the team and the work description. When developing artifacts or devices, the looks might be more important for a user than the functional features and sometimes it might be the other way round. This means that designers might be brought in first to shape the looks and the engineers later on to integrate the technology. On the other hand, it might even be the other way round where engineers work out the technical details and designers shape the looks afterwards.

An ideal situation would be, especially in a multi-, inter- or transdisciplinary team, when all involved members would participate in the design and development process from the beginning to the end. This would shape a collaborative atmosphere where designers and engineers could bring in their full potential. Others mentioned that designers should be constrained to limit unwanted innovation or creativity (Culverhouse 1995) This might be true when time is limited, but doesn't follow the general way of working as a designer, in which creativity and innovation is generally wanted and supported. In a transdisciplinary work it is important to guide and steer innovation in the right way and frame the work of designers so that designers know where creativity or rather innovative solutions are needed and where simpler or existing solutions might be better to use.

To provide such guidance, again, a well-managed team leadership is needed. This involves that structures in a team are given to support a close cooperation between designers and engineers. Methods like rapid prototyping or agile work methods that are getting more and more common to both engineers and designers present a large potential in facilitating collaborations in interdisciplinary teams.

As the question on how to support and nourish the teamwork is of utmost importance, when looking at interdisciplinary or transdisciplinary teams, it is important to have a well-defined work structure with procedures that allow for interwoven problem solving collaboration. This means to involve multiple or all team members to discuss issues and to decide solutions and features. At the same time, it is important to not underestimate the talents of individual team members. Therefore, individual tasks and rolls must be clearly defined and distributed according to the knowledge and skills in the team.

When looking back at our project, in which designers and engineers worked together to create a head-mounted projection display (HMDP), the situation got even more complicated when other stakeholders were involved. For our example, actors were selected as users for the HMDP to support and rehearse their performance. This meant that designers and engineers needed to work with actors who are artists and have a very different way of working and thinking compared to designers and engineers. This new composition of project members allowed for new possibilities. Generally, new views, ways of thinking and expert knowledge from a targeted profession of the designed artifact is beneficial to integrate into a team and to use as a source of knowledge.

Actors as potential users of a designed artifact are of course of central importance in a user-centric design. To understand the needs of actors and how to design and develop for their specific work environment, without spending large amounts of time for background research and gaining such knowledge was essential. Actors involved in an interdisciplinary team, could not only provide user experiences with the HMPD but also expert knowledge and valuable ideas and anticipated solutions in a collaborative design and development process.

Having an interdisciplinary team has its benefits in diversity of knowledge and ideas but needs interwoven structures and connections in order to facilitate understanding and efficient work among team members. Both universities and industry have begun looking for T-shaped engineers conceived by David Guest in 1991 as "a variation on Renaissance Man" with both deep and broad competences (Guest 1991) and researchers (IFM 2008)(Leonard-Barton 1995)(Palmer 1990) but it seems to be an early stage when it comes to exploring transdisciplinarity as a way to handle interdisciplinary teams successfully. We see transdisciplinarity as an interesting way of solving interdisciplinary problems but see that trained personal and researchers need to be found that can lead such interdisciplinary teams. We have mentioned before that transdisciplinarity teams are not new and have their challenges, such as disagreeing on methodologies that should be used to research or work on a certain topics. Nonetheless, we also see a large potential in interdisciplinary teams consisting in T-shaped researchers and engineers that are led by skilled transdisciplinary leaders.

### 5.2 Transdisciplinary research for automotive multi-operating system environments

As practical exploitation of information is rapidly pervading all spheres of society including technology, more and more of control processes are delegated to information processing devices (computers) and control applied in automotive industry started to transform from classical mechanics-based to information and computation based control. Over the past 30 years cars have changed from purely electro-mechanical vehicles to increasingly complex computerized systems through introduction of more complex features. This is not only driven by the necessity to provide innovations that improve sale rates, but also driven by customer demands and advances in technology. In 2014 new features were up to 70% software related (Bosch 2014). Those can be categorized into different domains: driver assistance (e.g. distance checking, lane assist), comfort (e.g. entertainment, navigation, communication) and safety related features (e.g. ASP, ESP).

Car manufacturers use over 100 years of experience and knowledge from mechanical and electrical engineering and about 30 years of experience in embedded software. 30 years ago software was rarely used in cars and the first electronic control units (ECU) were independently used for dedicated basic tasks (Broy et al. 2007). Today, a basic car architecture includes up to 100 ECUs (Ebert & Jones 2009), which are interconnected through a sophisticated communication infrastructure. Finally, an interface to the driver, the human-machine-interface (HMI), provides access to up to 700 functions of a car (e.g. (BMW 2014)). The overall development of cars is multi-disciplinary and transdisciplinary. Many teams work on different types of problems and contribute to the subsystems- as well as overall product/functions of a car. Introducing new technologies constantly increases the amount of disciplines in this process. Thus (Winner 2013) states, that "today systems are virtually impossible to develop within one engineering discipline" and relates to a manifold of necessary disciplines.

This is reflected in the internal organisation of car manufacturers, where all departments are created and separated based on their own discipline: mechanical engineering, electronics, ergonomics, etc. Software engineering is most often a subdivision of electronics. Departments produce modules, which are integrated in a common car platform (Pötsch 2011). Additionally, complex tasks (e.g. subsystems or specialized components) are often outsourced and commissioned to suppliers. Departments and suppliers rely on the concept of modularity, i.e. the exchange of strict sets of requirements and interface descriptions. However, modular development requires contextual knowledge to interconnected parts. Contrary to earlier expectations of complexity outsourcing, (Cabigiosu et al. 2013) states that "modular design does not substitute for high-power interorganizational coordination mechanisms". This supports the current change, especially in research and development departments, towards interdisciplinary and transdisciplinary departments, such as e.g. concept development, which consists of transdisciplinary teams from the fields of design, ergonomics and psychology.

Guidelines for user-centred development of a Driver Assistance System (DAS) (König 2016) state, that "a proven development strategy is to use an interdisciplinary team (human engineering team)". It further describes that the members of this team "must at the very least include engineers and psychologists". But, why are psychologists supposed to be part of the team? In order to understand the answer, the interconnection of the DAS component to other components has to be known. The DAS is connected to the HMI, which is used by the driver. "Physiology and traffic psychology is necessary in order to take into account the demands and the behaviour of drivers" (König 2016). This knowledge contributes to the behaviour and functionality of the DAS.

There is a shift from multidisciplinary to interdisciplinary to transdisciplinary research and development. The more interconnections exist, the more knowledge is necessary to solve particular problems. Automotive components are highly interconnected with other components, systems and services and it is inevitable to understand all of their implications. However, working in multi- or inter-disciplinary teams may cause developers to automatically gain enough knowledge to call the development trans-disciplinary – depending on the level of integration. The gap between inter-disciplinary and trans-disciplinary collaboration is very narrow.

Taking an illustrative application example from the automotive industry, where software developers have to implement prototypes of a new digital instrument cluster (such as digital speed/ rpm gauges as seen in e.g. (Audi 2014), (Audi 2016)). This software will be part of the HMI used in a real test-car in order to conduct user studies and to prove or disprove certain factors of the instrument cluster. In order to implement this particular piece of software, a developer has to understand the technical parts of a problem, e.g. how to obtain the speed information from the cars bus system, how to implement safety critical software on a certain platform and how to use a graphical processing unit (GPU) and related frameworks to implement the graphical part of the application. The actual concepts, related story boards and designs are usually created by designers, psychologists and/or ergonomists. However, to implement those in software, the developer must be able to understand the material. For example, how to convert/transform graphical artefacts or certain file formats into a usable piece of code in software. If special user interactions, such as gestures, are required, the developer also has to implement algorithms to detect those gestures. Thus in this specific example software developer has to collaborate with hardware developers, embedded software developers, designers and ergonomists and synthesize information/knowledge from variety of knowledge domains such as computer science, design and ergonomics in a transdisciplinary manner. Mono-disciplinarity in this context provides only the starting ground from which a collaborative project develops.

In our research about automotive multi operating system (Multi-OS) environments every researcher works in a different interdisciplinary field, which depends on the component or layer they are working on. Automotive Multi-OS environments compose multiple heterogeneous electronic control units (ECUs) to single ECUs in order to reduce the amount of hardware, wiring and weight in a car, consequently lowering production costs. This is possible through new technology, which provides powerful hardware and features for hardware/software virtualization. It allows multiple operating systems (OS) to be executed concurrently on a single hardware platform, i.e. an ECU. However, a composition of multiple ECUs or devices is a difficult task and causes problems and challenges on different categorized layers. Lower layers constrain higher layers, while higher layer depend on lower layers (Holstein et al. 2015).

A change of hardware in the lowest layer might cause the user-interface (UI) to be unusable. A concrete example is the change of touch screen size and resolution. A bigger screen size might be difficult to use while driving (Rümelin & Butz 2013) and a higher resolution might texts to be displayed too small and thus difficult to read

(Stevens et al. 2002). Both changes might have a negative effect on user-experience, which is part of the UI layer. Automotive environments have strong requirements regarding safety- and security-related software. Therefore, interconnections between certain components might be restricted or limited. In case of Multi-OS environments an OS is confined to its own hardware resources and only certain interconnections between the different OSs are allowed (Holstein & Wietzke 2015). A lower layer is responsible for the security/safety mechanisms and the transfer of data from one OS to another. An OS may have access to internet services or app stores. This means there is a risk of malicious third party software or faulty software, which in case of an error would only affect a single OS.

In the previous example transdisciplinary knowledge has been used in development. The latter example of our research shows a more profound usage of trans-disciplinarity. Here, trans-disciplinarity will help to understand the implications of inter-connections between different parts of a complex system: How do changes in certain layers affect the overall software architecture? How do restrictions and constraints in lower layers affect the development of user interfaces? Is the separation of operating systems through virtualization leading to a more secure architecture, besides the fact, that a homogeneous user interface requires the previously separated parts to be interconnected? In certain projects a transdisciplinary approach might be essential to the outcome and success of the project.

**5.3 Applied Interaction Design Research – a transdisciplinary practice. Examples from the information-intense industrial machinery**

*Designing products concerns applying technology in a form that brings usefulness and value to the user. As software takes an increasing portion of the product development, the more advanced products can be made while still providing a good experience to the user. In this section we will exemplify how transdisciplinary teamwork enhance research in product development, using examples from industry and the connection between industry and academia, arguing that it is even more beneficial the more we enter the infosphere.*

Design of interactive digital systems concerns forming an interaction between human users and the artefacts used by them. Design in this perspective is so much more than visual form and esthetics, though they are important components. Design thinking refers to cognitive activities used when designing symbolic and visual communication, material objects, activities and organized services, and complex systems or environments for living, working, playing, and learning (Buchanan 1992).

Kapor defines design in the Software Design Manifesto: "What is design? It's where you stand with a foot in two worlds - the world of technology and the world of people and humans purposes - and you try to bring the two together." (Kapor 1991). The above definition indicates that practicing interaction design is a field involving the application of knowledge, from domains outside of its own field. It acquires input from areas such as human behavior and psychology, from art as well as more traditional design fields, such as architecture and typography. To be able to design for the tasks performed, there must also be knowledge about the application area where the task is performed, the technology to apply and the surrounding eco-system.

**Researching Creation of products for the infosphere generation**

Since the above definitions where made, the impact of interaction design has increased a lot in the intercommunication between humans and the digital domain, such as the buzz and commercial success factor identified with "user experience" (Kuutti 2009). As software based systems are getting increasingly complex, the availability of computing power, sensors and actuators made software a much more integrated part of many products and systems, thus providing more and more of the functionality and value. In the automotive space, up to 70% of all innovation in products is currently software related (Bosch 2014). Even though not all of this innovation is related to interaction design, the way the user interacts, both in a sense of receiving information and being able to control the device, can be imperative for the user experience and safe operation of the device. The designer needs to understand how the technology works in combination with the user. This evolution of software systems impacting our life will likely continue, for example with the Internet of Things, where innovations move even further into the era of information generation and information processing and creating what (Floridi et al. 2010), (Floridi 2010) call "infosphere" that is informational environment corresponding to "biosphere".

Creating systems and products that collect, make use of information and provides a applicable and comprehensible result requires an interdisciplinary approach between natural sciences, social and human sciences and systems theory (Hofkirchner 2013). Such as, that the processes and the real world need to be understood in order to synthesize them into models of computation (Wallmyr 2015). Furthermore it involves the integration from information architectures and means of communication, to technical engineering and functional aspects of getting the different pieces of the system working together. In this, the interaction designer's role is to make sense of these systems and applications to the user. It is important that designers understand the application, technology and theory in order to successfully apply the generalized methods and principles of user interaction. The interaction designer needs to possess T-shaped competence (Guest 1991) (Boehm & Mobasser 2015) which means deep knowledge of at least one field and working knowledge of the current problem domains that makes it possible to bridge different research fields and approach technical issues when building utility for the user. Through understanding of different fields a common ground is found that facilitates improved collaboration and result. As Lindell argues, the interplay between interaction design and software engineering is problematic as these two activities have different epistemology. But treating information and code as a material can bring the two traditional disciplines into a combined craftsmanship (Lindell 2014).

Building custom solutions from scratch is many times not a viable option; instead development is done as integration of sub-parts with necessary adaption. Industrial products need to sustain the sometimes harsh environment, be sturdy to withstand years of tough usage, integrate with the way of working, and comply with market standards for e.g. emission and resistance. The final product might then be in production for ten years, with a subsequent lifetime of decades where service and replacement part is needed. Combining these criteria make it nearly impossible use parts from standard of-the-shelf or consumer market and in many cases the single industry domain cannot handle the investment and development themselves. Such an example is the forestry-harvesting sector, where new types of interfaces, using head-up-displays, have the potential to increase harvesting efficiency. However, their market alone is too small to support developing of the technology (Löfgren et al. 2007). Investments are thus distributed, as products are done in layers of existing components, software and technology. They must be generic enough so that several markets can use the same product.

To successfully select the right parts and build the right product for many markets needs a team that incorporate and exchange knowledge from several areas, not only technical but also on different market needs. Product realization project often requires a mix of different disciplines, such as mechanical designers, electric engineers, software developers, purchasers, production representatives, prototype builders etc. these teams are often lead by one or several roles, such as project managers, product managers or scrum masters. This creates multidisciplinary teams where different professions work together to build a product.

**An industrial example of transdisciplinary research into infosphere construction**

The question is: Is it necessary to have transdisciplinary teams in order to build next generation products? Perhaps, as working only interdisciplinary the project is constrained by the different professions focusing on their respective problems and solutions. This can lead to increased integration work, more late adjustments and a final outcome that does not reflect the bigger picture. In the case of a company developing hardware and software for industrial machinery it would mean clear disadvantages.

We have studied a company, that was going from a sub-supplier role to creating products of their own design, which in this specific case was a display computer. The company had several years of experience in building custom hardware and software. However, in retrospective it became evident that when building our own product we over focused on our in-house disciplines, our key knowledge that was normally the key contribution in customer projects. Functionality was added because it was technically possible, like TV, radio, modem and GPS. However, the market was either not interested or mature enough to appreciate it, thus leading to an overly complicated and expensive unit. One factor was that the development was not working interdisciplinary between electrical engineering, purchasing and software. Electric components were for example selected that did not have proper driver support, for the chosen operating system. Leading to massive efforts in integration of software and hardware. Another example was the industrial design that only covered mechanic design. The interaction with the display became much more of an office computer experience than what users in industrial machinery where normally accustomed to.

To continue the product development case, following display generations showed higher levels of interdisciplinary work. Such examples where electronic design and software design decisions made much more transdisciplinary, resulting in better component choices and easier integration. Also, a better understanding of customer needs lead to a hardware-software integration layer in software, making it possible for customers to move between different product families with minimal adaptation of their added application software. Simultaneously, the new industrial designer could incorporate both mechanical and software design, leading to a much more coherent experience for the end user. Other disciplines involved were production providing knowledge and experience on efficient production and service.

The above case illustrates how a higher degree of involvement and interaction between disciplines during product design and realization can result in a more integrated, thought-out and purposely facetted product. This is however not the only purpose of the illustration. The other argument is that transdisciplinarity is something that evolves continuously, (Dorothy Leonard-Barton 1995) as the knowledge transfers between disciplines and individuals within the project. One of enablers of this process is continuous design review where different team members not only review the current solution, but also exchange knowledge on the factors in technology, usage, cost etc. that contribute to the solution made. As a result, experienced teams that have more interaction with other professions, become more transdisciplinary integrating tighter with other disciplines, understanding their vocabulary, limitations and possibilities. At a managerial level transfer can also be facilitated to share information and create new contacts, for example through information exchange events or relocation of personnel. As an example we can mention Canon that relocates its research and development center every six month (Harryson 1997).

As mentioned, many industries that were earlier more focused towards mechanization and automation are now progressing into the infosphere. More understanding is needed how to efficiently use all this information to benefit users in their respective domains. One factor is to avoid information overload, in automotive system to the level of awareness needed in vehicle interaction solutions, such as for visual perception given in (Wördenweber, B., Wallaschek et al. 2007), p. 48 that explains how vision constructs reality for an observer. Apart from perception and awareness, another aspect is making the information accessible to the user in an understandable and attractive manner. An example of an industrial application sector of interest is agriculture. Here information technology provides a vital piece when addressing how to efficiently and sustainably to produce food to a growing population. However a farmer or a machine operator is by profession neither a computer professional nor an information analyst. Thus, the move into more information based production systems centered on software engineers' preferences might be associated with obstacles. (Sørensen et al. 2010) mention that even though the use of computers and internet have improved acquiring of external information as well as management and processing of internal information, "the acquisition and analysis of information still proves a demanding task". The availability of data does not warrant the understanding or usefulness of the data to the user (Chinthammit et al. 2014).

Supporting the transformation into information driven applications thus calls for more transdisciplinary development that will provide connection among variety of technologies and between technology and the user, society and environment. To connect to the prior case, we are coming closer to what can efficiently be interacted with using normal displays. Future development, for example see-through interfaces that augment reality and interfaces that use more-than-human visual perception to exchange information. This development has to include competences in information architecture and information design, data communication not only with local system but also cloud communication as well as haptic interaction with the user. In addition even deeper knowledge of industry domain is necessary to build the information system and computation models that provide more automation as well the right information to the user.

**6. Bridges between academia and industry – education and the industrial PhD**

Another aspect of transdisciplinarity is the connection between research and industry where one of many methods transdisciplinary bridges between academia and industry is built by industrial PhD students. Giving a dual direction transfer where the industrial researcher brings real-world research questions from industrial settings into academia, while simultaneously bringing information and results from the research community into industry. This ongoing exchange builds information exchange contact points as well as basic understanding of different fields, through persons that can bridge different disciplines and domains of knowledge production.

Industrial projects are to a large extent limited by a fixed description of requirements and defined task to realize, within given resource limits, such as initial time estimation. Thus these projects cannot in the same way as research elaborate on different ways to address problems and solutions. Instead the industrial PhD can bring findings and results from academia into the industrial projects, building on the research findings and adding the needed parts to design and develop or improve an industrial product.

On the other side of the bridge, the research side, it is instead encouraged to seek new and novel solutions. As researchers we are encouraged to publish our results, making information and findings from our work available to the wider research community. Going to conferences and otherwise seeking information for our own research, give inspiration and input when observing results from other application domains, thus sharing and receiving information.

Interaction design research has though been criticized for counteracting its own purpose, with the argument that design science should not be about a *science of design* but rather a *science for design*. Instead of being bound by past research it should instead be free to critically examine and question results of scientific research, with the aim of envisioning the future (Krippendorff 2007). Simultaneously design research has been questioned for not valuing application of the science to a specific field as a research result and valid contribution (Chilana et al. 2015), thus perhaps limiting the possibilities for interaction design researchers to endeavor into interdisciplinary research. It can be argued that in order to foster interdisciplinary research and improved collaboration, possibilities should be offered that would enable such efforts and results to be published. At present, transdisciplinary research still meets difficulty to find its proper place in academia, that is traditionally organized by disciplines, and publications are by far and large purely disciplinary with a confined view of what constitutes a good contribution. The trend of subdivision of classical disciplines into ever more narrow sub-disciplines should be counteracted by the synthetic approaches of transdisciplinarity that bring cohesion into the otherwise completely disconnected islands of knowledge. As scientists we work often with understanding of the world and how we can improve it. As such, a fitting conclusion is to refer back to Krippendorff: "*Design concerns what could work in the future, a future that is more interesting than what we know today.*"

## 6. Conclusions

One of the central issues of transdisciplinary knowledge production is communication of information and knowledge across the disciplinary and cultural borders. How do we interpret the same object (boundary object) from the perspective of different disciplines, expressed in their domain languages?

How can our research which ranges from interdisciplinarity to transdisciplinarity contribute both to the existing knowledge as well as further development of practice and theory connecting information, communication, computation, (inter-)action and cognition? Let us examine how some of the characteristics of transdisciplinarity reflect in our research.

**Reflexivity**
Reflexivity relates both to the inner relationships between knowledge domains as well as the self-reflection over the proposed problem solution and its meaning for the stakeholders. It implies asking both the questions *why* and *how*, that makes relation to value systems and ethical deliberation important. The bottom line of every decision-making is the value system (which affects how we see our goals) and the sense of (feeling and understanding of) the current state of the world, that is understanding of where we are and where we want to go.

**Epistemic and value-basis transparency**
Ethical and epistemic conceptualizations are closely coupled. (Tuana 2015) Epistemic transparency in the research project requires insight in one owns assumptions and knowledge-related choices. Visibility of value grounds, decision-making transparency and analysis are central to the success of a transdisciplinary project. Coupled ethical-epistemic analysis has helped in the past projects identify new and refined research topics, and informed modeling for multi-objective, robust decision-making. (Singh et al. 2015) One of important attitudes in knowledge generation over several epistemic domains is attentiveness and respect for both knowledge and ignorance granted for all stakeholders Uncertainties and inadequate knowledge play should be identified and carefully tackled.

**Addressing the complex architecture of the knowledge space**
Understanding of the complex architecture of the multi-level and multi-dimensional knowledge space is a part of reflexive relation to knowledge production. The roles of various stakeholders must be well understood and benevolent mutual communication based on shared goals secured. For example, in medicine, addressing problem of disease requires understanding processes from molecular to cellular and level of organs, the whole organisms and their environment, including psycho-social factors thus knowledge in such a transdisciplinary project is a result of a synthesis and derivation of knowledge from all those classical academic domains in conjunction with its "users" medical institutions, societal groups, etc. In the case of our HCI field- designers, developers, users and other stakeholders are involved in the process of knowledge production.

**Syntactic vs. semantics vs. pragmatic aspects of knowledge**
As research operates on different levels of knowledge production, all three layers of semiotics are involved: syntactic – often coding of a computer program, semantics – design of programs and other artifacts for specific purposes and pragmatics – study of the behavior of the artifact (design) in practice – use-case studies.

**Integration process**
Important part of the transdisciplinary process is integration, which presents ontological and epistemological as well as organizational challenges. Riegler (Riegler 2005) mentions the following types of problems met in transdisciplinary integration process: (P1) unfamiliarity among different disciplines with a mutual information deficit; (P2) different terminology – different use of the same terms; (P3) different aims of scientific work – prediction vs explanation; (P4) hard sciences vs. soft sciences; (P5) Basic research vs. applied science and (P6) Individual vs. group research. Riegler addresses this topic from the point of view of constructivist approach which is interested in how exactly different contributions can be integrated in a common framework. He emphasizes the importance of the *common worldview*, the minimum shared commitment that makes it possible to relate different positions, and identify differences and similarities, granularity level of knowledge and other characteristics. Experiences from our projects indicate as well that commitment must be shared in order for a project to succeed.

**Embodiment and embeddedness of information, computation, cognition, communication**
One important aspect of research that has an ambition to have relevance for the real life is embodiment and embeddedness, which brings the element of sensualizing. Instead of abstract ideas of solutions, applied research deals with embodied problems that bring sensory qualities to the technological solutions and makes esthetic aspects of design necessary to address. Here we meet decisions made based on function (including its ethical aspects) vs. esthetics and experiential dimensions. Human – computer interaction (HCI) design does not only describe or contemplate possible futures – it builds concrete artefacts that *set material constraints* on our possible futures. At this stage ICT have applications even in arts and artistic production. (Busch 2009)

**Questions we want answers to**
Info-computational approaches today are in the center of our contemporary knowledge production. Both in literal sense of ICT used to communicate information and compute knowledge as well as in a sense of models based on information and computation. Answers fundamental questions: What is reality? What is life? Why do things happen? What is intelligence, mind, and understanding? What will happen next? Why does anything even exist? – are given in terms of infocomputation, providing new and more understandable answers than ever before. Ordinary people can nowadays "see" what atoms do, how quarks behave, how galaxies collide, how universe evolves since the big bang or what possible consequences of global warming might be – all of it via visualizations of computer simulation results. In not so distant future we will have similar possibilities to see alternative consequences of our possible political, economic and other choices. It will bring whole new possibilities for democratic decision-making. As humans we are interested not only in how things are, but we also want to know what we can expect, what is possible and what are the consequences. The promise of new theories, discoveries, inventions and developments will be possible to study in ever increasing detail and in much more systematic and multifaceted, multidimensional way. We want to know **why** *and we want to act based on deep understanding that takes into account not only logic, but the totality of human experience.*

**The underlying logic of change**
Most often logic is taken as tacit part of the theory construction, frameworks for reasoning and action. However, it should be noticed that logic is a research field on its own and a fast developing too. Nicolescu addressed the question of logic especially with regard to the axiom of excluded middle, (Nicolescu 2010) which states that nothing can be at the same time A and non A. This axiom reflects the interest of Aristotle and ancient Greeks in general an interest that is still predominant to this day in structures that *persist*, and not in the process of *change*. Transdisciplinarity on the other hand is centered on change. It means that the dynamics of process, when a structure is partly in the current and partly in the next state, is vital, where the middle is necessary included, if the process is continuous. Differences in logics imply differences in what can be expressed and argued for and how. One interesting approach in the context of transdisciplinary research is Brenner's Logic in reality (Brenner 2008) especially applied to the dynamics of information. (Brenner 2012)
Mathematician Chaitin-Chatelin argues in her book *Qualitative computing* that the Aristotle's classical logic is too limited to capture the dynamics of nonlinear computation. As the necessary tool for addressing the nonlinear dynamics she proposes the organic logic. This logic will be the core of the "Mathematics for Life" yet to be developed (Chaitin-Chatelin 2012). Yet another logical development was Zadeh's fuzzy logic where the "excluded middle" was replaced with a spectrum of possibilities. Even though classical logic is widely used and considered adequate, for better understanding of the process of knowledge production, integration and synthesis, logics that put their focus on dynamical process and nonlinearity are of great interest.

**Addressing the issue of learning in transdisciplinary research projects**
The aim of research is traditionally not only to solve concrete problems, but also to contribute to the learning, that is to the shared knowledge of the community of practice (research community, knowledge building community, culture). However, being often focused on specific and real-life problems, transdisciplinary research faces the problem of comparative analysis of the research findings of different groups with different approaches, different stakeholders, values and preferences. Should transdisciplinary research be seen as an alternative to free deliberation such as commonly used in political, social or business decision-making, based on common sense and personal experience of stakeholders? Or can it be used to contribute to the development of classical research fields by informing them about the real world context in which abstract frameworks and academic discourses can be placed in? This two-way learning process can be obtained through individuals who belong both to research in the Mode 1 and the Mode 2 (Gibbons et al. 1994), such as industrial PhD students as we described. The interdisciplinary and transdisciplinary knowledge production is new in the university world with its long and persistent traditions, and the best way to contribute to better understanding of the applicability and the role of different modes of knowledge production is to educate future generations of researchers and citizens not only in disciplinary research methods but also in interdisciplinary and transdisciplinary research.